\documentclass{article}

\usepackage{graphicx}
\usepackage{natbib}
\usepackage{epsfig}
\usepackage{psfrag}
\usepackage{fancyhdr}

\usepackage{color}
\definecolor{gray}{gray}{.5}

\usepackage{bbding}

\usepackage{amsmath,amssymb}
\usepackage{amsbsy}

\title{\bf Formation mechanism of hairpin vortices in the wake of a truncated square cylinder in a duct}

\begin{document}

\author{Vincent DOUSSET and Alban POTH\'{E}RAT\\%
Applied Mathematics Research Centre, Faculty of Engineering and Computing,\\
Coventry University, Priory Street, Coventry CV1 5FB, United Kingdom}

\date{10 February 2010}

\maketitle

\cfoot{}
\renewcommand{\headrulewidth}{0pt}

\begin{abstract}
We investigate the laminar shedding of hairpin vortices in the wake of a truncated square cylinder placed in a duct, for Reynolds numbers around the critical threshold of the onset of vortex shedding.
We single out the formation mechanism of the hairpin vortices by means of a detailed analysis of the flow patterns in the steady regime.
We show that unlike in previous studies of similar structures, the dynamics of the hairpin vortices is entwined with that of the counter-rotating pair of streamwise vortices, which we found to be generated in the bottom part of the near wake (these are usually referred to as \emph{base vortices}).
In particular, once the hairpin structure is released, the base vortices attach to it, forming its legs, so these are streamwise, and not spanwise as previously observed in unconfined wakes or behind cylinders of lower aspect ratios.
We also single out a trail of $\Omega$-shaped vortices, generated between successive hairpin vortices through a mechanism that is analogous to that active in near-wall turbulence.
Finally, we show how the dynamics of the structures we identified determine the evolution of the drag coefficients and Strouhal numbers when the Reynolds number varies.
\end{abstract}

%\pacs{Valid PACS appear here}

%\begin{keywords}
%Truncated square cylinder wake, Hairpin vortex, Laminar wake
%\end{keywords}

\section{Introduction}

Flows past a truncated cylinder are involved in a wide variety of problems dealing with the design of e.g. buildings, ship funnels or chimney stacks.
The typical values of the Reynolds number $Re$ considered in these problems are greater than $10^4$.
In contrast, though relevant with probes used in intrusive measurement methods \citep{bhm08} and heat transfer in electronic circuit boards \citep{ni04}, the flow dynamics at lower $Re$, {\it i.e.} up to $10^3$, have been scarcely investigated.

In these regimes, four main structures are generated in the cylinder wake: the horseshoe system, the free shear layer stretching from the cylinder free end, those stretching from both cylinder lateral faces and a system of trailing vortices.
By convenience, we assume that the cylinder is mounted on the bottom wall.

The horseshoe pattern results from the interaction between the boundary layers arising respectively at the cylinder upstream face and at the bottom wall \citep{baker79,lhd08}.
It is formed by a system of swirls generated by flow separation at the cylinder front that spirals around the cylinder.
It remains steady for $Re<1500$ without any restriction on the thickness of the wall boundary layer \citep{lhd08}.

The system of trailing vortices consists of two pairs of counter-rotating streamwise vortices located below the cylinder tip and above the cylinder base, respectively.
According to their location, they are referred to as {\it tip} and {\it base vortices}.
In both cases, their origin still remains an open question.
From experimental flow visualisations in the wake of a truncated circular cylinder at $Re>10^4$, \cite{ef76} and \cite{pl00} suggested that the tip vortices were generated above the upper cylinder face from the rolling-up of the lateral ends of the upper free shear layer, while \cite{khhmk84} interpreted them as resulting from the tilting of the lateral free shear layers in the vicinity of the cylinder free end.
From experimental measurements in the same configuration at $Re>10^4$, \cite{shd04} suggested that the tip vortices had nothing to do with the lateral free shear layers.
Investigating experimentally the flow past a truncated square cylinder for $200<Re<10^4$, \cite{wz09} drew the same conclusions as \cite{khhmk84}.
On the other hand, little information is available on the base vortices.
\cite{ef76}, \cite{shd04} and \cite{wz09} agree that they result from the tilting of the lateral free shear layers in the vicinity of the bottom wall.
\cite{ef76} furthermore indicated that the base vortices were initially aligned along the spanwise direction and then tilted along the streamwise axis between the mid-span and the free end.
The cylinder aspect ratio $\gamma$, defined as the ratio of the cylinder height $h$ to the characteristic length of its cross-section $d$, determines which of the tip or the base vortices prevail in the wake \citep{shd04,wz09}.
Also, increasing the thickness of the boundary layer at the bottom wall strengthens the base vortices \citep{wzcl06}.

Since the horseshoe structure remains steady for $Re<1500$, any vortex shedding occurring in this regime is fed by the free shear layers only and the cylinder aspect ratio $\gamma$ determines whether the vortex shedding is asymmetric, symmetric or absent.
For high values of $\gamma$, the vortex shedding is governed only by the lateral free shear layers and an asymmetric K\'arm\'an-like vortex street is observed \citep{lsc05,sa83,wz09}.
For intermediate values, the transition to unsteadiness leads to a symmetric vortex street formed by hairpin vortices aligned on the wake centreline \citep{sa83,sho87}.
Finally, small values of $\gamma$ prevent any kind of vortex shedding \citep{khhmk84,ni04}.

The critical values of $\gamma$ separating these regimes depend on the shape of the cylinder cross-section.
The latter affects the shape of the free shear layers as well as the position of the wake structures with respect to one another and thereby the mode of vortex shedding \citep{sa83,sho87}.
Also, the thickness of the boundary layer at the bottom wall has a noticeable influence on the determination of the critical values of $\gamma$ \citep{sa83,khhmk84}.

The formation mechanism of the K\'arm\'an vortices relies on the alternate roll-up and shedding of the lateral free shear layers as in non-truncated cylinder wakes.
In contrast, there has not been yet any clear agreement on the scenario leading to the symmetric hairpin vortex street.
From experimental investigations, \cite{sa83} suggested that both lateral free shear layers joined the top one to form a single entity in the near-wake and as the latter became unstable, an arch-type vortex was formed and released in the wake.
This view is supported by \cite{hy04} who performed numerical simulations of the flow past a truncated square cylinder with $\gamma=0.5$ for $Re\le2000$ and by experimental flow visualisations by \cite{wz09} at $Re=221$.
The latter authors also included the tip and base vortices within the arch-vortices released in the wake.
\cite{yik07} simulated the flow past a truncated square cylinder featuring $\gamma=1$ at $Re=500$ and claimed that hairpin vortices were originally vortices that detached from the free shear layer stretching from the cylinder free end and then grew into hairpin vortices.
Until now though, the generation mechanism of hairpin vortices in the wake of a truncated cylinder has never been the object of any dedicated study so it remains rather unclear.

In this article, we investigate the wake of a truncated cylinder of square cross-section using three-dimensional direct numerical simulations.
Our goal is to clarify the dynamics of the formation of hairpin vortices.
Our approach relies on a detailed analysis of the flow structures in the steady regime.
We restrict our investigations to cylinder wakes confined in a duct of rectangular cross-section, in which the inlet velocity profile is a well defined paraboloid flow.
Under these conditions, the scenario we shall bring forward is new and differs from those inferred in previous works in several points.
In addition, we have detected a set of $\Omega$-shaped vortices forming between two successive hairpin vortices.
We shall also give the details of the mechanism of their formation.
The present flow configuration and the numerical set-up are introduced in section \ref{sectionConfiguration}.
In section \ref{sectionSteadyRegime}, we identify the steady flow patterns and their dynamics throughout the steady flow regime.
Section \ref{sectionUnsteadyRegime} is dedicated to the description of the formation and release of hairpin vortices at the onset of vortex shedding.
The formation of the $\Omega$-shaped vortices and the effects of increasing $Re$ within the unsteady regime are also detailed in this section.
Finally, in section \ref{sectionFlowCoefficients}, we show how the evolution of the flow coefficients is affected by the flow  dynamics and, in particular, how the appearance of secondary recirculation regions on the cylinder lateral faces noticeably affects the viscous drag coefficient.

\section{Configuration and numerical set-up \label{sectionConfiguration}}
\subsection{Configuration}

\begin{figure}%[h]
\psfrag{af}{(a)}
\psfrag{bf}{(b)}
\begin{center}
\psfrag{b}{$2b$}
\psfrag{a}{$2a$}
\psfrag{d}{$d$}
\psfrag{h}{$h$}
\psfrag{u}{${\bf U}$}
\psfrag{x}{$x$}
\psfrag{y}{$y$}
\psfrag{z}{$z$}
\includegraphics[width=0.95\textwidth]{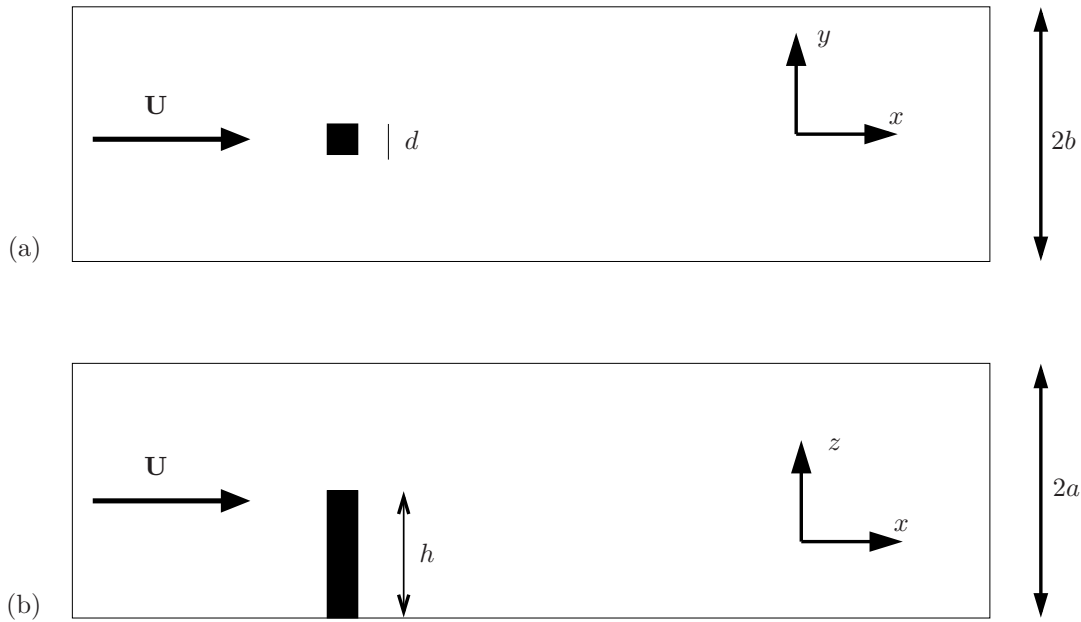}
\caption{Configuration of the study: (a) top view; (b) side view.}
\label{configuration}
\end{center}
\end{figure}

We consider the flow of an incompressible fluid (density $\rho$, kinematic viscosity $\nu$) past a truncated square cylinder in a rectilinear duct of rectangular cross-section as shown in figure \ref{configuration}.
A square cylinder (width $d$ and height $h$) is mounted on the duct bottom wall at equal distance from both duct side walls.
The upstream cylinder face is normal to the streamwise direction taken along the $x-$axis.
The cylinder axis is parallel to the $z-$axis.
The origin of the frame of reference is located at the centre of the upper cylinder face.
The duct height ({\it resp.} width) along the $z-$axis ({\it resp.} $y-$axis) is $2a$ ({\it resp.} $2b$).
The present configuration features $h=4d$, $2a=8d$ and $2b=10d$, which yields a transverse ({\it resp.} spanwise) blockage ratio $\beta_t=d/2b=0.1$ ({\it resp.} $\beta_s=h/2a=1/2$) and a cylinder aspect ratio $\gamma=h/d=4$.
$\beta_t$ is low so that it does not have any noticeable effect on the horseshoe system \citep{bbf98}, while the effects of the spanwise flow confinement shall be discussed throughout this study.

\subsection{Flow equations and numerical set-up \label{sectionFlowEquationsNumericalSetup}}

We study the flow using three-dimensional direct numerical simulations with a finite-volume code based on the OpenFOAM 1.4.1 framework \citep{wtjf98}.
The flow motion is governed by the Navier-Stokes equations (\ref{massequation}) and (\ref{momentumequation}), written in their non-dimensional form as:
\begin{eqnarray}
{\bf \nabla} \cdot {\bf u} & = & 0 \label{massequation} \\
\frac{\partial {\bf u}}{\partial t} + ({\bf u} \cdot {\bf \nabla}) {\bf u} & = & \frac{1}{Re} \nabla^2 {\bf u} - {\bf \nabla} p, \label{momentumequation}
\end{eqnarray}
where ${\bf u}$ and $p$ are the velocity and the pressure fields, respectively, and $Re=U_0d/\nu$ ($U_0$ is the maximum velocity at the inlet).

The numerical domain is $15d$ long upstream and $30d$ long downstream the cylinder.
A no-slip impermeable boundary condition is imposed at all walls through a homogeneous Dirichlet condition and a homogeneous Neumann condition is applied for the velocity at the outlet.
This geometry ensures that the outlet boundary has no noticeable feedback onto the cylinder wake as it is located more than $25d$ away from the cylinder \citep{snd98}.
Preliminary numerical computations at a prescribed $Re$ are performed in the duct with the cylinder absent.
The resulting velocity profile of the fully-established flow in the cross-section is used as the inlet boundary condition for the velocity in the three-dimensional simulations with the cylinder present.
For the pressure field, a homogeneous Neumann condition is imposed at all boundaries but at the outlet where a homogeneous Dirichlet condition is applied.
The flow equations (\ref{massequation}) and (\ref{momentumequation}) are solved in a segregated way and the PISO algorithm as detailed in \cite{wtjf98} is used to deal with the pressure-velocity coupling.
Second-order discretisation using central difference schemes in space and a quadratic backward scheme in time has been used on a non-uniform Cartesian mesh.

The numerical code has been successfully tested on the two-dimensional flow past a square cylinder at $Re=150$ \citep{doolan09} and by simulating the three-dimensional flow past a non-truncated square cylinder at $Re=200$ as described in \cite{mgmb00}.
The latter test has been performed on a non-uniform mesh featuring $1.4\times 10^6$ points, {\it i.e.} slightly less points than in \cite{mgmb00}.
The non-dimensional minimum grid size was set to 0.03 so that a better resolution of the boundary layer was achieved than in \cite{mgmb00}, whose computations were performed on a fully uniform mesh.
We have observed mode-A flow patterns together with dislocated K\'arm\'an vortices as in \cite{mgmb00}.
Our computed values of the Strouhal coefficient $St=fd/U_0=0.164$ and the pressure part of the drag coefficient $C_{Dp}=2F_D/(\rho U_0^2 hd)=1.70$ ($f$ is the vortex shedding frequency and $F_D$ the force obtained by the integration of the pressure field over the cylinder surface) have recovered the values of \cite{mgmb00} within 1\% and 3\% respectively, which ensures that the present numerical code addresses the dynamics of three-dimensional cylinder wakes within a fully satisfactory accuracy.
\begin{table*}
\center
      \begin{tabular}{p{4.5cm}ccc}
      Meshes         & M1 & M2 & M3 \\ \hline%\hline
	Number of nodes along ${\bf e}_z$ & 32 & 64 & 96 \\
        $\delta_z$ & 0.06 & 0.03 & 0.02 \\
      Total number of nodes & $7.4\times10^5$ & $1.5\times10^6$ & $2.2\times10^6$ \\
      $\epsilon_{st} = |1-St(Mi)/St(M3)|$ & 0.10 & 0.03 & / \\
      $\epsilon_{cd} = |1-C_D(Mi)/C_D(M3)|$ & $3.51\times10^{-3}$ & $3.05\times10^{-3}$ & / \\ \hline%\hline
      \end{tabular}
      \caption{Main characteristics of the different meshes and errors in drag coefficient $C_D$ and Strouhal number $St$ relative to M3 mesh at $Re=200$.
$\delta_z$ is the non-dimensional distance between the upper cylinder face and the grid point nearest to the latter face.
All meshes feature 135 and 120 non-uniformally distributed nodes along the $x$- and $y$-axes, respectively.}
      \label{charmesh}
\end{table*}

We have then implemented a test to estimate the required number of mesh nodes for the simulations with the truncated cylinder wake.
The main characteristics of the tested meshes are provided in table \ref{charmesh}.
In the plane normal to the cylinder axis, we have taken over the same mesh as that used in the test on the non-truncated cylinder wake which have yielded very satisfactory results.
We have then varied the number of points along the $z-$axis and the distance $\delta_z$ between the cylinder top face and the grid point nearest to it.
We have simulated the flow at $Re=200$ over 1000 turnover times $t_u=d/U_0$ and derived the total drag coefficient $C_D=2F_x/(\rho U_0^2 hd)$ ($F_x$ is the streamwise component of the total force exerted by the flow on the cylinder) and the Strouhal number $St$ in each case.
We have found that both the errors in $C_D$ and $St$ relative to mesh M3 decreased with the number of nodes, which shows good convergence.
In order to save CPU time and keep a reasonable accuracy in our computations, we shall perform all our simulations with M2 mesh.

A final validation test has been performed on the configuration of \cite{lsc05}.
The authors investigated the flow past a truncated circular cylinder with aspect ratio $\gamma=10$ between two infinite parallel impermeable walls placed at the cylinder bottom end and at 5 cylinder diameters above the cylinder free end.
We have simulated the flow at $Re=100$ using the same geometry, computational domain and boundary conditions.
For this test, we have designed a structured mesh consisting of a polar mesh embedded in a square of two cylinder diameter width and a rectangular Cartesian grid covering the rest of the computational domain.
The mesh contains about $1.8\times10^6$ nodes with a non-dimensional minimum grid size of 0.02 between the cylinder upper face and the grid point nearest to it.
We have computed the Strouhal number at $Re=100$ and found $St=0.1431$ which represents an error of $1.34\%$ compared to the value of $St=0.145$ in \cite{lsc05}.
In addition, our simulations have recovered a uniform spanwise distribution of the Strouhal number along the cylinder span in agreement also with \cite{lsc05}.

In summary, our code has been successfully validated on two different configurations of flow past a truncated cylinder involving either a transverse or a spanwise flow confinement.
In both cases the unsteady flow dynamics have been accurately captured by the computation of the Strouhal number.
We shall now consider the truncated cylinder wake in a rectangular duct in which the flow is confined along both transverse and spanwise directions.

\subsection{Vortex identification}

To explain the formation mechanism of hairpin vortices, we need to identify flow structures and, in particular, vortices.
Establishing an objective definition of a vortex has been the subject of several studies in the literature.
\cite{jh95} have assessed different approaches of identification and tracking of vortical structures.
One of them is based on the analysis of the eigenvalues of the symmetric tensor ${\bf S}^2 + {\bf \Omega}^2$, where ${\bf S}$ and ${\bf \Omega}$ are the respective symmetric and antisymmetric part of the velocity gradient tensor ${\bf \nabla u}$.
In this approach, a vortex core corresponds to a pressure minimum not induced by viscous effects nor unsteady straining.
It is defined as a connected region with two negative eigenvalues of ${\bf S}^2 + {\bf \Omega}^2$.
A vortex is therefore detected at a given location in the fluid domain if the median eigenvalue, denoted $\lambda_2$, is locally negative.
This approach is particularly efficient at spotting ring-type vortices \citep{jh95}.
Unfortunately, it delivers no information on the rotation directions of the vortical structures.
For this reason, flow patterns will also be characterised by iso-surfaces of vorticity, especially in the unsteady flow regime.
In the steady one, we will analyse the streamlines, since they match the flow trajectories \citep{cpc90}. 

\section{Steady flow regime \label{sectionSteadyRegime}}

We have performed nine successive simulations at increasing $Re$ for $10\le Re\le400$.
Unsteady flows have been computed over a total simulation time close to $1000t_u$.
Our computations have yielded a steady flow regime for $Re\le150$ and an unsteady regime for $Re\ge200$.
In no case does the resulting flow feature any closed streamlines, in line with the findings of \cite{hapw78}.

\subsection{Flow patterns at $Re=100$}
\begin{figure}%[h]
\psfrag{z1}{$h_1$}
\psfrag{z2}{$h_2$}
\psfrag{z3}{$h_3$}
\psfrag{xu}{$x_u$}
\psfrag{xd}{$x_d$}
\psfrag{s1}{$S_1$}
\psfrag{s2}{$S_2$}
\psfrag{s3}{$S_4$}
\psfrag{s4}{$S_3$}
\psfrag{hs}{HS}
\psfrag{hv}{HV}
\psfrag{F1}{$F_1$}
\psfrag{F2}{$F_2$}
\psfrag{x}{$x$}
\psfrag{y}{$y$}
\psfrag{z}{$z$}
\psfrag{2a}{(a)}
\psfrag{2b}{(b)}
\psfrag{2c}{(c)}
\psfrag{2d}{(d)}
\center
\includegraphics[width=0.99\textwidth]{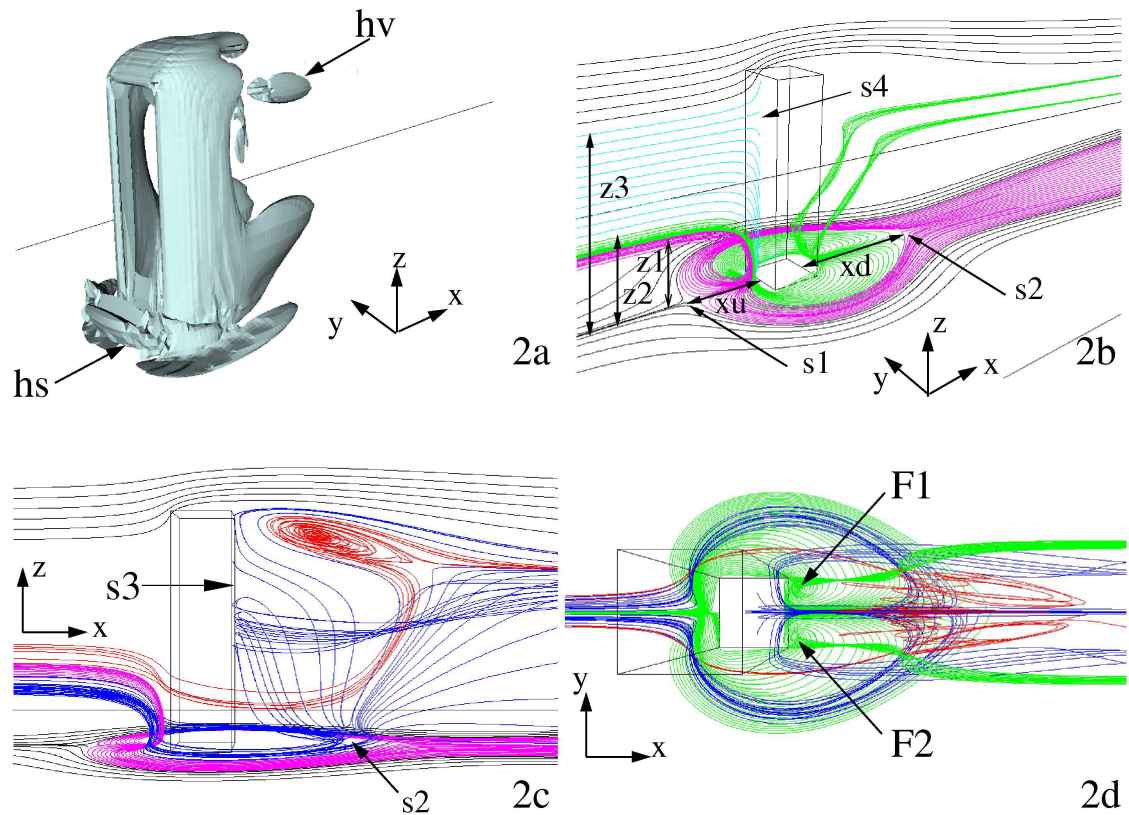}
\caption{Steady flow patterns at $Re=100$. Iso-surfaces of $\lambda_2=-0.05$ (a) and kinematic streamlines with horseshoe pattern in pink, trailing vortices in green, streamlines rejoining the stagnation points on the upstream cylinder face in cyan, head vortex in red, spanwise vortices in blue, black streamlines are only deflected by the cylinder wake without noticeably influencing it: three-dimensional (b), side (c) and top (d) views.}
\label{flowdyn}
\end{figure}

Figure \ref{flowdyn}(a) shows the steady flow patterns at $Re=100$ depicted by iso-surfaces of $\lambda_2=-0.05$, {\it i.e.} less than $0.2\%$ of the absolute minimum in $\lambda_2$, and by kinematic streamlines in figures \ref{flowdyn}(b-d).

In figures \ref{flowdyn}(b-c), one sees that the black streamlines that are simply deflected around the wake structure separate from those that enter it at saddle point $S_1$ located at $x=-1/2-x_u$ in the centre-plane ($y=0$).
All streamlines other than the black ones enter the wake structure.

\subsubsection{Streamlines originating in the upstream centre-plane}
Pink and green streamlines originate in the upstream centre-plane ($y=0$), follow a similar path around the cylinder and separate at saddle point $S_2$ located at $x=1/2+x_d$ in the centre-plane ($y=0$) \citep{hapw78,shsy03}.
Pink lines depict the horseshoe pattern as shown in experiments by e.g. \cite{lhd08}.
The front horseshoe swirl, denoted HS in figure \ref{flowdyn}(a), is observed upstream of the cylinder base at $x=-1$.
The green set of streamlines impacts the bottom duct wall downstream of a nodal point of attachment $N_a$ just in front of the upstream cylinder face, as observed in \cite{hy04}.
It then recirculates behind the cylinder within the horseshoe pattern where it reaches two foci, denoted $F_1$ and $F_2$ in figure \ref{flowdyn}(d), and subsequently spirals upwards and eventually tilts along the $x-$axis to create a pair of counter-rotating streamwise vortices.
This corresponds to the base vortices identified by \cite{ef76} and \cite{shsy03}.
The foci were also detected in numerical simulations by \cite{hy04} and \cite{yik07}, but were not linked to the formation of base vortices.
At $Re=100$, these vortices are very weak so that they only induce a weak pressure minimum that is difficult to capture with the iso-surfaces of $\lambda_2$.
Far upstream, in the centre-plane $(y=0)$, the border between the pink and green sets of streamlines is located at $z=h_1-\gamma$ and the upper border of the green set at $z=h_2-\gamma$.

Interestingly, we did not isolate any streamwise tip vortices, whether in the iso-surfaces of $\lambda_2$ or in kinematic streamlines.
There are two probable reasons for this.
Firstly, the spanwise flow confinement imposes a strong spanwise velocity gradient similar to that induced in thick boundary layers and \cite{wzcl06} showed that the thickening of the bottom boundary layer enhanced the base vortices at the expense of the tip ones.
Secondly, the present cylinder aspect ratio is rather low and no tip vortices have been observed in simulations run with a low aspect ratio \citep{shsy03,yik07}.

A third set of streamlines (in cyan) originating within the centre-plane $(y=0)$ just above the green set of streamlines at ($h_2-\gamma \le z \le h_3 - \gamma$) ends at a line of front stagnation points on the upstream cylinder face along the wake centreline.
A half-saddle point $S_3$, indicated in figure \ref{flowdyn}(b), is detected along this line of stagnation points slightly below the cylinder tip.

\subsubsection{Streamlines rejoining the centre-plane downstream}
Blue and red streamlines rejoin the centre-plane $(y=0)$ downstream of the cylinder.
Blue streamlines originate upstream, below those generating the horseshoe pattern [see figure \ref{flowdyn}(c)] but slightly off the wake centreline [see figure \ref{flowdyn}(d)].
On the one hand, the streamlines located just upstream of $S_2$ head upwards and back upstream along the wake centre-plane behind the cylinder and separate into two substreams at half-saddle point $S_4$ on the downstream cylinder face [see figure \ref{flowdyn}(c)].
The lower substream is deflected by this downstream face outwards the wake along the $y$-axis and eventually carried away by the free stream to form a pair of counter-rotating spanwise vortices.
The upper substream heads towards the cylinder tip where the free stream takes it away along the wake centreline.
On the other hand, the blue streamlines located just downstream of $S_2$ head upwards and downstream until they reach the free stream where they tilt along the $x-$axis.

A gap between the blue streamlines moving upstream and downstream is observed at the rear of the cylinder slightly below the cylinder tip where a transverse vortex is seen [red lines in figure \ref{flowdyn}(c-d)].
This head vortex consists of a pair of symmetric transverse vortices located at short distance off the wake centreline and connected to a single transverse vortex at the wake centreline.
Only the right part of the head vortex, denoted HV, is visible on the $\lambda_2$-iso-surfaces from figure \ref{flowdyn}(a).
The streamlines enter this three-vortex structure at the periphery of the symmetric vortices, spiral towards the axis and exit at the periphery of the centre vortex onto the wake centreline.
Note that the head vortex is not generated by streamlines curling from the cylinder free end as in \cite{hapw78}, \cite{sg81} and \cite{wz09}, but from streamlines circulating upwards from the cylinder bottom half as in \cite{shsy03} and \cite{yik07}.

Throughout this description of the flow topology at the bottom duct wall and cylinder faces at $Re=100$, we have isolated three nodal points ($N_a$, $F_1$, $F_2$), two saddle points ($S_1$, $S_2$) and two half-saddle ones ($S_3$, $S_4$).
This topology thus satisfies the mathematical criterion established by \cite{hapw78}.

Finally, two free shear layers arise from the lateral downstream edges of the cylinder and stretch on both sides of the wake (see $z-$vorticity contours in figure \ref{vortexStreetRe200}).
Due to the presence of the bottom duct wall and the free shear layer stretching from the top trailing edge of the cylinder, their streamwise length is reduced at both their ends and reaches its maximum in the vicinity of the base vortices.
We insist that all the steady flow patterns detected in the present computations are generated exclusively by streamlines engulfed under these lateral free shear layers.

\subsection{Effect of increasing $Re$ within the steady flow regime}

As $Re$ is increased within the steady flow regime, the sets of streamlines rejoining the line of stagnation points at the cylinder upper face extends downwards, while the upstream spanwise extension of the set of streamlines generating the base vortices (green lines) shrinks and that feeding the horseshoe pattern (pink lines) broadens.
The increase of $Re$ therefore implies an increase of $h_1$ and a shrinking of both $h_2$ and $h_3$.

The overall shape of the horseshoe pattern however hardly changes: the main swirl HS [see figure \ref{flowdyn}(a)] drifts slightly upstream as observed also in \cite{baker79} \cite{hy04} and \cite{shsy03}, but $S_2$ barely moves, {\it i.e.} $x_u$ increases and $x_d$ changes little.
The base vortices gain in strength and entangle with the spanwise vortices in the vicinity of the foci $F_1$ and $F_2$, which get closer to the rear cylinder face as $Re$ is increased.

At $Re=150$, secondary recirculation regions appear on both lateral and top faces of the cylinder.
The lateral recirculations are made of stretched and twisted blue streamlines and the top one results from flow separation at the cylinder top face as in \cite{yik07}.
The top recirculation deflects the outer stream further up, which, combined with the increased strength of the head vortex, results in the latter shifting into a position higher than the cylinder tip in the late stages of the steady flow regime.
In this regime, the head vortex also moves farther downstream.
Finally, at $Re=150$, the pair of symmetric vortices forming the head vortex merges into the centre-plane $(y=0)$ so that the streamlines are taken away downstream once they come out of the eye of this single remaining structure.

\section{Unsteady flow regime \label{sectionUnsteadyRegime}}
\subsection{Description and formation mechanism of the symmetric vortex street at $Re=200$}

\begin{figure}
\psfrag{a3}{(a)}
\psfrag{b3}{(b)}
\psfrag{c3}{(c)}
\psfrag{d3}{(d)}
\psfrag{e3}{(e)}
\psfrag{f3}{(f)}
\psfrag{af}{}
\psfrag{bf}{}
\psfrag{cf}{}
\psfrag{df}{}
\psfrag{ef}{}
\psfrag{ff}{}
\psfrag{x}{$x$}
\psfrag{y}{$y$}
\psfrag{z}{$z$}
\psfrag{TV}{TV}
\psfrag{RV}{RV}
\psfrag{LV}{LV}
\psfrag{SL}{SL}
\psfrag{SR}{SR}
\psfrag{BL}{BL}
\psfrag{BR}{BR}
\center
\includegraphics[width=0.9\textwidth]{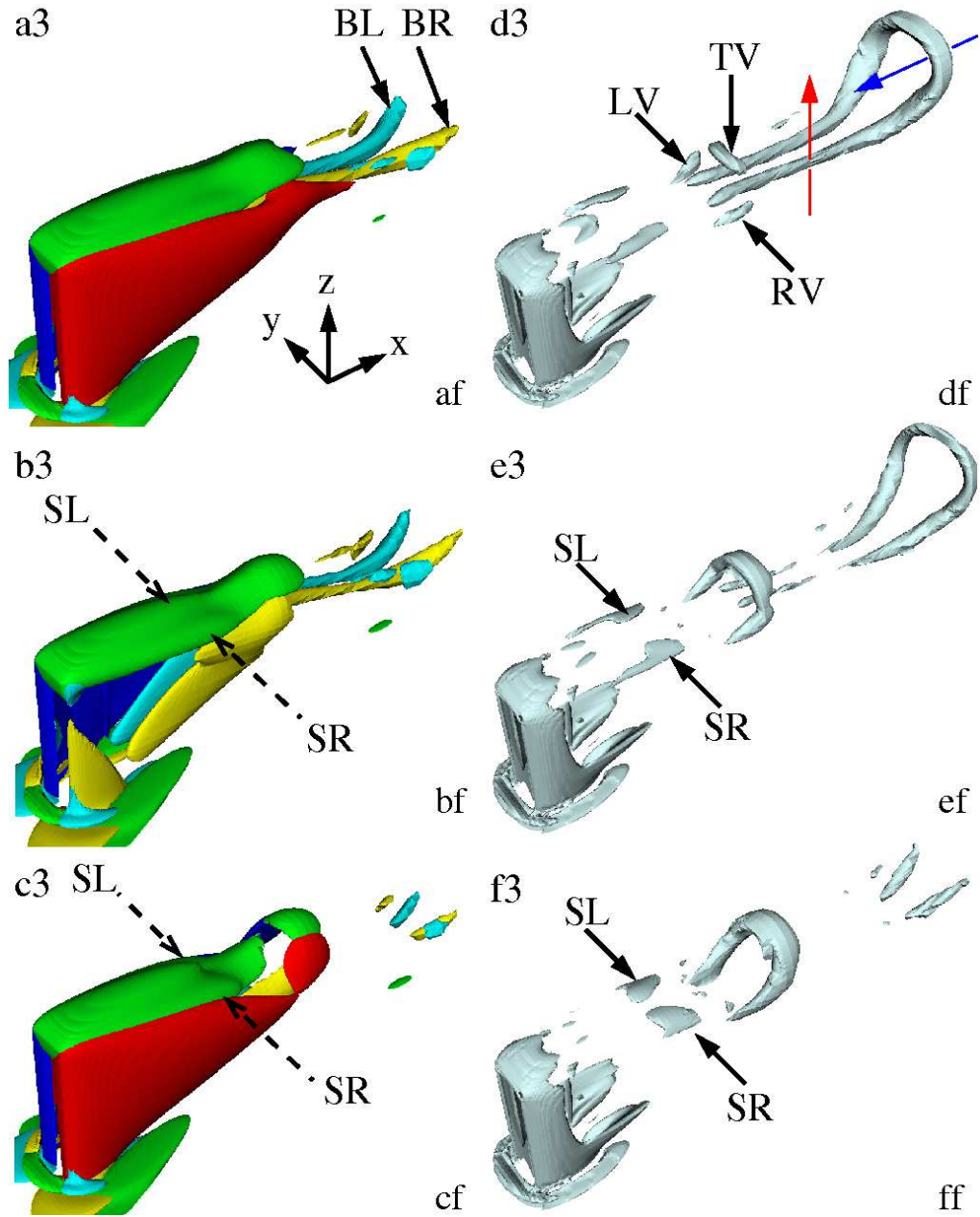}
\caption{Vortex street at $Re=200$ depicted by iso-surfaces of non-dimensional vorticity ${\bf \omega}$ (a-c) and $\lambda_2=-0.5$ (d-f) at $t=1349t_u$ (top), $1351t_u$ (middle) and $1353t_u$ (bottom).
$\omega_x=2.5$ ({\it resp.} $\omega_x=-2.5$) in yellow ({\it resp.} cyan); $\omega_y=5$ in green; $\omega_z=5$ ({\it resp.} $\omega_z=-5$) in red ({\it resp.} blue).
The red lateral free shear layer is not shown in (b).
TV, LV and RV are vortical structures shed from the top, left and right free shear layers, respectively.
BR and BL are the base vortices.
SR and SL are the secondary streamwise vortices clearly visible on $\lambda_2$-iso-surfaces but not on vorticity iso-surfaces where their locations are indicated by dashed arrows.
Red and blue arrows on (d) indicate the jets induced by the base and hairpin vortices respectively, whose interaction participates to the generation of $\Omega$-shaped vortices.
Supplementary movie 1 (available at journals.cambridge.org/flm) shows an animation of the vortex street at $Re=200$.}
\label{vortexStreetRe200}
\end{figure}

At $Re=200$, unsteadiness appears through a periodic laminar symmetric shedding of hairpin vortices aligned on a single row along the wake centreline.
This vortex shedding has been observed in experiments by \cite{sa83} and \cite{sho87} and numerical simulations by \cite{hy04} and \cite{yik07}.
The formation and release of hairpin vortices depicted by iso-surfaces of vorticity field and of $\lambda_2=-0.5$ ($0.5\%$ of its absolute minimum) are represented at three successive time instants during the formation and release of the hairpin vortices at $Re=200$ in figures \ref{vortexStreetRe200}(a-f).

At $t=1349t_u$, one observes a hairpin vortex in the early stage of its formation in the near cylinder wake and a freshly released, fully formed, hairpin vortex in the far wake.
Three distinct vortical structures denoted TV, LV and RV, respectively originating from the top, left and right free shear layers, are singled out in figure \ref{vortexStreetRe200}(d).
TV indirectly results from the mutual interaction of the base vortices BR and BL, shown in figure \ref{vortexStreetRe200}(a), that generates an upwards jet between them and lifts the tail of the top free shear layer.
This results in an inversion of the curvature of the latter which becomes unstably curved \citep{liou94}, and leads to the breakaway of its tail (TV).
At $t=1351t_u$, two secondary counter-rotating streamwise vortices, denoted SR and SL in figures \ref{vortexStreetRe200}(b-f), then rise just upstream of the structure shed from the top free shear layer and wrap around the base vortices, while TV, LV and RV have gathered into a single bow of vorticity.
The subsequent pairing between both the base and the secondary streamwise vortices eventually triggers the symmetric shedding of structures LV and RV from both lateral free shear layers [blue and red iso-surfaces in figure \ref{vortexStreetRe200}(c)] observed at $t=1351t_u$.
At this moment, the head of the hairpin vortex is completely formed and taken away by the free stream, while the base vortices detach to join the head of the hairpin vortex and form its legs visible on the hairpin vortex present in the far wake in figure \ref{vortexStreetRe200}(d).

Our computations show that the hairpin vortices are not generated only by the destabilization of the top free shear layer as suggested in \cite{hy04} and \cite{yik07}.
Instead, they result from the smooth assembling of structures shed from the initially steady flow patterns.
The head of the hairpin vortex is formed from the smooth assembling of vortical structures detached from the top and lateral free shear layers.
The hairpin vortices observed in the present simulations differ from the arch-type vortices detected in experiments by \cite{sa83} and \cite{wz09}.
Firstly, due to the spanwise velocity gradient imposed by our flow confinement, the hairpin head is located at the downstream end of the structure.
The free stream velocity is higher in the vicinity of the cylinder tip so that the hairpin head, located in this region, is carried away further downstream than the rest of the hairpin structure.
Secondly, the hairpin legs are almost parallel to the streamwise axis and not aligned along the spanwise direction as those of the arch-type vortices.
In our configuration, the hairpin legs are indeed the base vortices which are oriented along the streamwise axis, whereas those of the arch-type vortices are the spanwise vortices shed from the lateral free shear layers.
Thirdly, in our configuration, neither the legs nor the spanwise vortical part of the hairpin vortex grow  {\it a posteriori} from the structure shed from the top free shear layer, as claimed in \cite{yik07}.
Instead, both these structures are present in the very first steps of the formation of the hairpin vortices.
This shows that the base vortices play an active part in the vortex shedding mechanism as in the flow past a trapezoidal tab \citep{dm04,yms01}, but unlike in \cite{sa83} and \cite{wz09}.
In the last two studies though, the inlet velocity is almost spanwise invariant and the shedding results from the interaction between the top and lateral free shear layers only.

\subsection{Formation and release of secondary $\Omega$-shaped vortices}

We have described how the hairpin vortices are generated and released in the cylinder wake.
In the unsteady regime, the structures that take part in their formation were detected by plotting iso-surfaces of vorticity and $\lambda_2$ for reasonably high values of these quantities.
We shall now refine our analysis and track weaker structures by bringing the threshold values of $\lambda_2$ and of the vorticity closer to zero for the flow at $Re=200$.

\begin{figure}
\psfrag{af}{(a)}
\psfrag{bf}{(b)}
\psfrag{cf}{(c)}
\psfrag{df}{(d)}
\psfrag{x}{$x$}
\psfrag{y}{$y$}
\psfrag{z}{$z$}
\psfrag{A}{A}
\psfrag{B}{B}
\psfrag{C}{C}
\psfrag{H1}{$H_1$}
\psfrag{H2}{$H_2$}
\psfrag{O1}{$\Omega_1$}
\psfrag{O2}{$\Omega_2$}
\psfrag{O3}{$\Omega_3$}
\psfrag{SR}{SR}
\psfrag{SL}{SL}
\psfrag{BR}{BR}
\psfrag{BL}{BL}
\center
\includegraphics[width=0.95\textwidth]{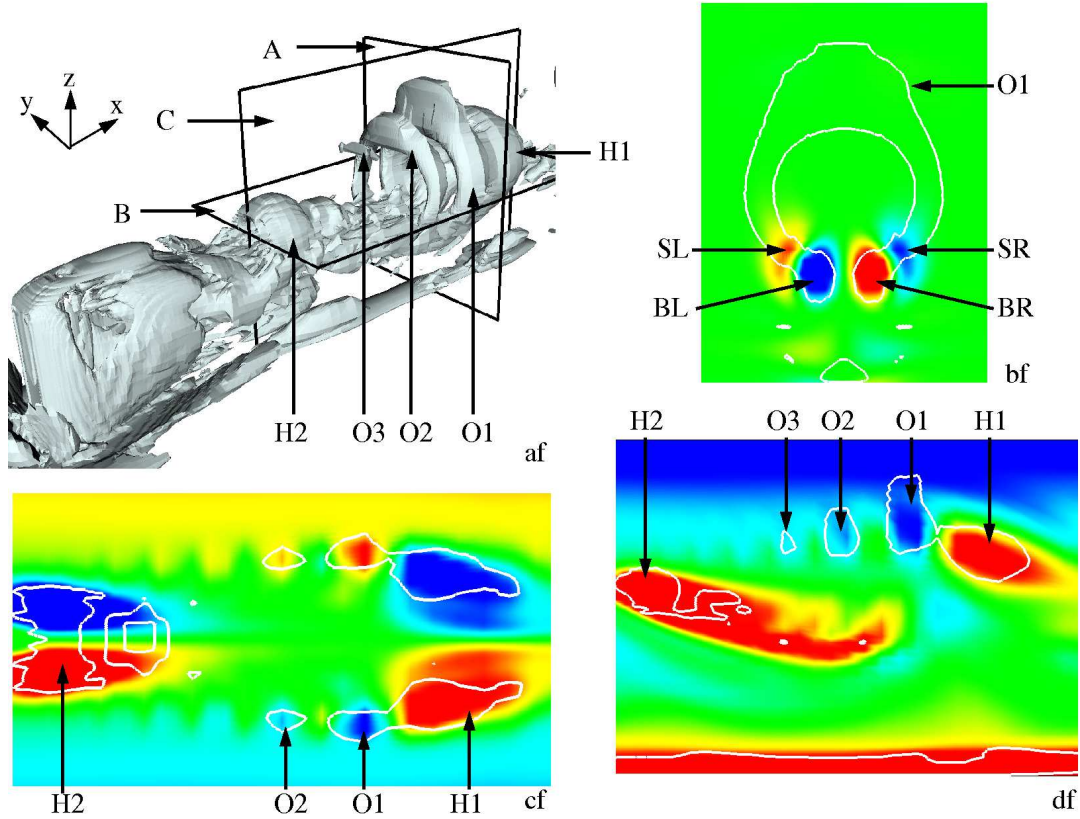}
\caption{(a) Iso-surfaces of $\lambda_2=-0.03$ at $Re=200$ and $t=1353$.
Fields of non-dimensional (b) $x$-vorticity in plane A ($x=12.9$) with $|\omega_x|\le2$, (c) $z-$vorticity in plane B ($z=0$) with $|\omega_z|\le1.5$ and (d) $y-$vorticity in plane C ($y=0$) with $|\omega_y|\le1.8$ (minimum and maximum values in blue and red, respectively).
Thick white lines on (b-d) are the intersected $\lambda_2-$iso-surfaces.
$H_1$ and $H_2$ are two successive hairpin vortices,
$\Omega_1$, $\Omega_2$ and $\Omega_3$ three $\Omega$-shaped vortices,
SR and SL the respective left and right secondary streamwise vortices 
and BR and BL the respective left and right base vortices (see also figure \ref{vortexStreetRe200}).}
\label{omegavortex}
\end{figure}

Figure \ref{omegavortex}(a) shows the three-dimensional vortex street at $t=1353t_u$ depicted by iso-surfaces of $\lambda_2=-0.03$ and two-dimensional vorticity iso-surfaces in three different planes are presented in figures \ref{omegavortex}(b-d).
One observes a chain of $\Omega$-shaped vortices between two successive hairpin vortices, denoted $H_1$ and $H_2$.
Three such $\Omega$-shaped vortices, denoted $\Omega_1$, $\Omega_2$ and $\Omega_3$, are detected in figure \ref{omegavortex}(a).
The strength of these vortices decreases with their distance to the downstream hairpin vortex.
They are located in $(y,z)$-planes and their bottom ends are connected to the secondary streamwise vortices, denoted SR and SL in figure \ref{omegavortex}(d).
The $\Omega$-shaped vortices are mirror images of the hairpin vortex as they all rotate in the sense opposite to it, as shown by figures \ref{omegavortex}(b-d).
Vortex streets involving hairpin vortices and secondary reverse vortices ({\it i.e.} rotating in the opposite sense to that of the hairpin head) have been observed in the flow past a trapezoidal tab in experiments \citep{yms01} and numerical simulations \citep{dm04}.
Reverse vortices detected in the latter configuration were however located below the line of hairpin vortices by contrast with our observations where hairpin and reverse vortices are all aligned.
The simulations by \cite{dm04} have yet clearly established the $\Omega$-shape of the reverse vortices as in our case.

The formation of $\Omega$-shaped vortices follows from a mechanism encountered in turbulent boundary layers in channel flows described in e.g. \cite{zabk99}.
During its downstream motion, the head of the hairpin vortex broadens and induces a backwards streamwise jet through the head [blue arrow in figure \ref{vortexStreetRe200}(d)].
The pair of streamwise vortices forming the legs of the hairpin vortex generates an upwards jet between them [red arrow in figure \ref{vortexStreetRe200}(d)].
The shearing between the latter jet and the one induced by the head of the hairpin vortex causes the formation of a bridging shear layer between both streamwise vortices.
This shear layer subsequently rolls up into an $\Omega$-shaped vortex in the ($y,z$)-plane.
This mechanism is active all along the legs of the hairpin vortex and leads to the formation of several $\Omega$-shaped vortices between two successive hairpin vortices.

Importantly, the generation of $\Omega$-shaped vortices is a direct consequence of the streamwise orientation of the hairpin legs.
This mechanism is thus absent in configurations where arch-type vortices have been observed \citep{sa83,wz09}.

\subsection{Effect of increasing $Re$ on the vortex street}

\begin{figure}
\psfrag{af}{(a)}
\psfrag{bf}{(b)}
\psfrag{x}{$x$}
\psfrag{y}{$y$}
\psfrag{z}{$z$}
\center
\includegraphics[width=0.85\textwidth]{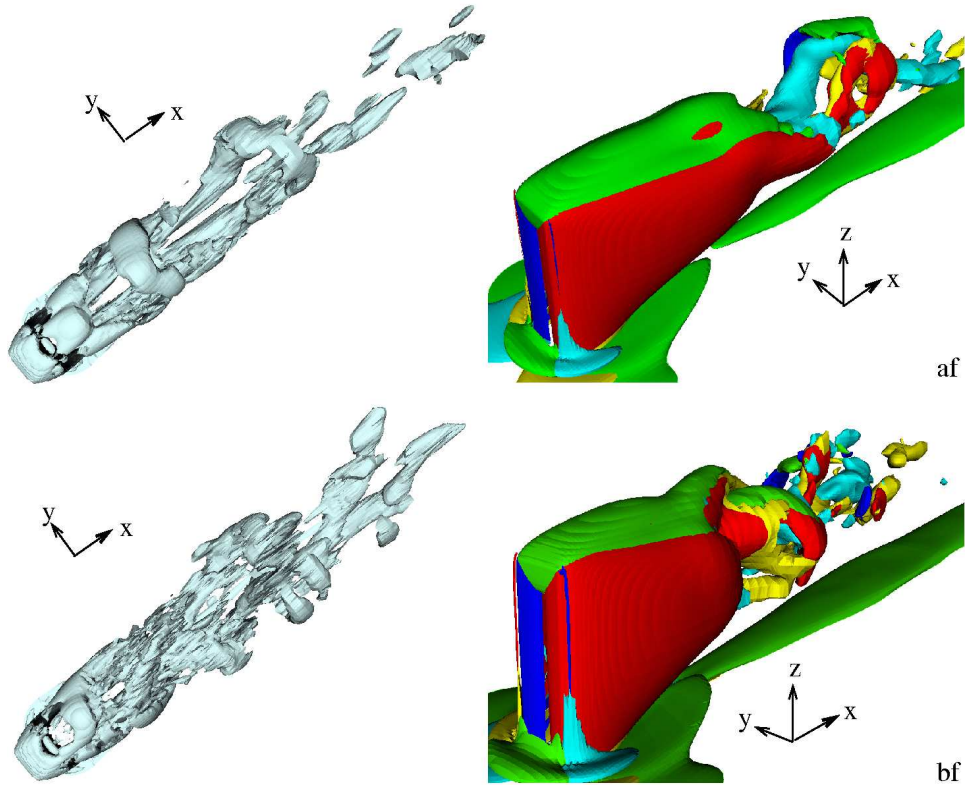}
\caption{Vortex street at $Re=250$ and $t=1410t_u$ (a) and $Re=300$ and $t=1044t_u$ (b).
({\it left}) Iso-surfaces of $\lambda_2=$ -0.5 (a) and -0.7 (b) and ({\it right}) fields of $\omega_x=\pm 3.75$, $\omega_y=6.25$ and $\omega_z=\pm6.25$ (a) and $\omega_x=\pm 4.25$, $\omega_y=9.25$ and $\omega_z=\pm 9.25$ (b) in yellow and cyan, green, red and blue, respectively.}
\label{irregularVortexStreetHa0}
\end{figure}

\begin{figure}
\psfrag{ttu}{$t/t_u$}
\psfrag{cd}{$C_D$}
\psfrag{re200}{$Re=200$}
\psfrag{re250}{$Re=250$}
\psfrag{re300}{$Re=300$}
\psfrag{re400}{$Re=400$}
\center
\includegraphics[width=0.65\textwidth]{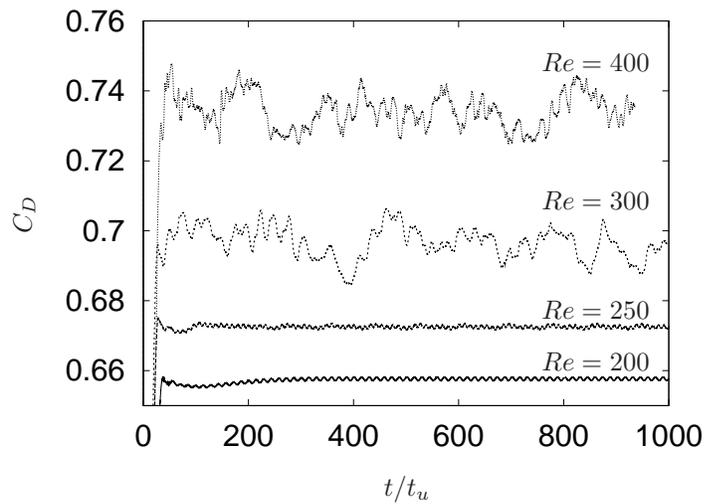}
\caption{Time histories of the total drag coefficient $C_D$ at $Re$=200, 250, 300 and 400.}
\label{cdtimeHistory}
\end{figure}

At $Re=250$, the mechanism of formation and release of hairpin vortices described previously remains unchanged, but the shape of the hairpin vortices becomes irregular and weak spanwise oscillations of the base vortices are observed.
The latter oscillations disturb the shedding of the top and lateral free shear layers so that the vortex street turns slightly asymmetric as shown in figure \ref{irregularVortexStreetHa0}(a).
For $Re\ge300$, these oscillations gain in intensity.
The legs of the hairpin vortices do not split at the same $z-$coordinate anymore, while their head turns into a chaotic aggregate of structures shed from the free shear layers.
Hairpin vortices are therefore released alternately on either side of the wake centreline and so a fully asymmetric vortex street is observed as seen in figure \ref{irregularVortexStreetHa0} (b).
The chain of $\Omega$-shaped vortices is still observed in this regime, but their shape is irregular too and their orientation follows that of the hairpin vortex that creates them.
A similar evolution of the vortex street with increasing $Re$ has been described by \cite{yyo93} in experiments of the flow past a square plate.
The gradual change from a regular to a chaotic vortex street for $200\le Re \le400$ is illustrated in figure \ref{cdtimeHistory} by the respective time histories of the total drag coefficient $C_D$, introduced in section \ref{sectionFlowEquationsNumericalSetup}.

\section{Flow coefficients \label{sectionFlowCoefficients}}
\subsection{Drag coefficients}
The variations of the total drag coefficient $C_D$ with $Re$ are presented in figure \ref{coefre}(a).
One observes that $C_D$ decreases within the steady flow regime and slightly increases in the unsteady one.
These variations reflect the dynamics of the vortex formation region as in the two-dimensional square cylinder wake \citep{se04}.
In steady regimes, it encompasses the pair of spanwise recirculation regions, while in unsteady ones it includes the formation region of the hairpin vortices.
In both cases, this region induces an adverse pressure gradient.
The increase of $Re$ within the steady regime results in the streamwise elongation of the symmetric steady recirculation regions.
This enhances the adverse pressure gradient which gradually counter-balances the streamwise pressure gradient induced by the duct flow and thus leads to the decrease in the total drag coefficient $C_D$.
In the unsteady flow regime, the mechanism is reversed as the formation region of the hairpin vortices shrinks, hence the increase in $C_D$.
Remarkably, the present values of $C_D$ obtained with a cylinder spanning over half the duct height are about half of those obtained with a non-truncated cylinder \citep{snd99}.

The viscous component of the drag coefficient is defined as $C_{Dv}=2F_v/(\rho U_0^2 dh)$ where $F_v$ is the viscous force acting on the cylinder.
We have reported its variations over the cylinder span at different $Re$ in figure \ref{coefre}(b).
One can identify three regions along the span: two narrow ones close to the respective cylinder ends and a larger one between them \citep{lsc05}.
$C_{Dv}$ exhibits peak values just where the flow around the cylinder is fast, {\it i.e.} below the cylinder free end and at short distance of the bottom duct wall within the horseshoe region.
In contrast, $C_{Dv}$ remains rather constant in the mid-span region where the flow recirculates much more slowly.
The corresponding values are therefore significantly smaller than the peak values reached in both end regions.
One observes also that $C_{Dv}$ turns negative for ($-0.4\le z/h \le -0.13$) at $Re=200$ and that the $z-$range with negative $C_{Dv}$ widens as $Re$ is increased.
This feature results from the appearance and subsequent widening of the secondary recirculation regions at the cylinder lateral faces for $Re\ge150$.
The associated counter-flow induces advert friction at the side walls that reduces the overall value of $C_{Dv}$ to the point of making it negative.
This effect was detected in two-dimensional square cylinder wakes by \cite{se04}.

\subsection{Spanwise lift coefficient}
We define the spanwise lift coefficient $C_z=2F_z/(\rho U_0^2 d^2)$ where $F_z$ is the $z-$component of the force exerted by the flow on the cylinder.
The evolution of $C_z$ computed only on the upper cylinder face versus $Re$ is reported in figure \ref{coefre}(c).
By definition, $F_z$ results from the integration of the pressure force on the cylinder upper face and is defined up to a constant determined by the value of the reference pressure.
As a consequence, only the variations of $C_z$ reflect the flow dynamics, while its absolute values shall be regarded relatively to the reference pressure.
In particular, the sign of $C_z$ bears no significance.
$C_z$ decreases in the steady regime and slightly increases in the unsteady one.
As for the drag coefficient, the decrease of $C_z$ results from the lengthening of the spanwise recirculation regions at the back of the cylinder.
The reason for its increase is the appearance of the secondary recirculation over the cylinder upper face that induces an adverse pressure gradient.
The destabilization of the top free shear layer in the unsteady regime for $Re\ge200$ slightly enhances this trend throughout the unsteady regime.

\subsection{Strouhal numbers}
\begin{figure}%[h]
\psfrag{af}{(a)}
\psfrag{bf}{(b)}
\psfrag{cf}{(c)}
\psfrag{df}{(d)}
\psfrag{cz}{$C_z$}
\psfrag{cd}{$C_D$}
\psfrag{st}{$St$}
\psfrag{re}{$Re$}
\psfrag{Reegala10}{$Re=10$}
\psfrag{Reegala100}{$Re=100$}
\psfrag{Reegala200}{$Re=200$}
\psfrag{Reegala300}{$Re=300$}
\psfrag{Reegala400}{$Re=400$}
\psfrag{zsurh}{$z/h$}
\psfrag{cdvnormalise}{$C_{Dv}/<C_{Dv}>_z$}
\psfrag{steady}{Steady}
\psfrag{instat}{Unsteady}
\psfrag{symmetric}{symmetric}
\psfrag{asymmetric}{asymmetric}
\center
\includegraphics[width=0.99\textwidth]{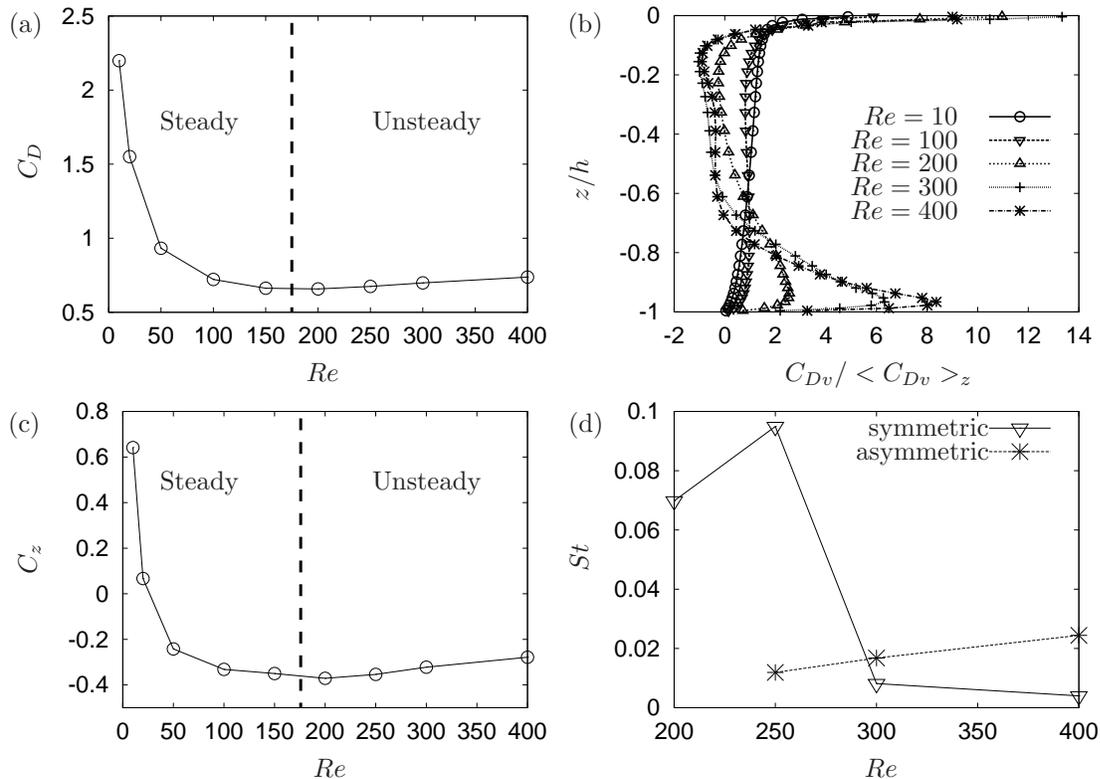}
\caption{(a) Total drag coefficient $C_D$ on the cylinder stem versus $Re$ and (b) spanwise distribution of the viscous drag coefficient $C_{Dv}$ at different $Re$ ($<C_{Dv}>_z$ is the $z-$average value and the time-average value of $C_{Dv}$ is used in the unsteady flow regime).
(c) Spanwise lift coefficient $C_z$ on the cylinder upper face and (d) Strouhal numbers $St$ associated to the symmetric and asymmetric modes versus $Re$.
The dashed line on (a) and (c) separates the steady and the unsteady flow regimes.}
\label{coefre}
\end{figure}

The frequency of the symmetric  ({\it resp.} asymmetric) mode was obtained from the time history of the total drag coefficient $C_D$  ({\it resp.} lift coefficient $C_L=2F_y/\rho U_0^2 dh$ where $F_y$ is the $y-$component of the force exerted by the flow on the cylinder).
We were then able to calculate the Strouhal numbers associated to the respective frequencies of the symmetric and asymmetric modes.
Their dependence on $Re$ is presented in figure \ref{coefre}(d).
At $Re=200$, the vortex street is indeed perfectly symmetric so the asymmetric mode is absent.
At $Re=250$, a slight asymmetry is observed in the wake due to the appearance of the small vertical oscillations of the base vortices.
These oscillations are enhanced when $Re$ is further increased until the wake becomes chaotic and completely asymmetric.
When the wake is symmetric at $Re=200$, only the symmetric mode is present with an associated $St=0.07$.
When the asymmetric mode appears at $Re=250$, its associated $St$ is much lower ($St=0.01$) but increases with $Re$.
As the wake becomes more asymmetric, the Strouhal number associated to the symmetric mode collapses down to values below those of the asymmetric one.

\section{Conclusions}

\begin{figure}
\psfrag{steady}{Steady}
\psfrag{unsteady}{Unsteady}
\psfrag{secondary}{recirculation}
\psfrag{recirculation}{at side and top faces}
\psfrag{symmetric}{symmetric}
\psfrag{asymmetric}{asymmetric}
\psfrag{shedding}{shedding}
\psfrag{10}{10}
\psfrag{150}{150}
\psfrag{200}{200}
\psfrag{250}{250}
\psfrag{300}{300}
\psfrag{400}{400}
\psfrag{re}{$Re$}
\center
\includegraphics[width=0.95\textwidth]{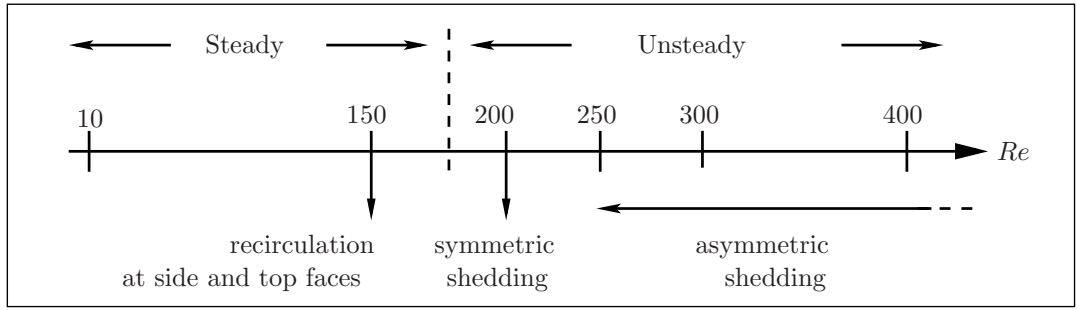}
\caption{Summary of the main flow features observed in the present simulations.}
\label{summarySimulations}
\end{figure}

We have investigated the flow past a truncated square cylinder in a duct of rectangular cross-section using three-dimensional direct numerical simulations for $Re\le400$.
Figure \ref{summarySimulations} summarises the main flow regimes we have encountered.
The steady regime has been thoroughly scrutinised and the flow patterns identified.
The present computations have shown that these patterns were all generated by sets of streamlines flowing around the cylinder foot and underneath the lateral free shear layers.
In particular, in our configuration, the spanwise flow confinement induces an important spanwise shear that promotes the development of the base vortices at the expense of the tip ones.
Also, we have singled out the appearance of secondary recirculation regions at the lateral and upper cylinder faces for $Re\ge150$.

At $Re=200$, the flow unsteadiness leads to a symmetric vortex street formed by a single row of hairpin vortices aligned on the wake centreline.
We have detailed a formation mechanism of the hairpin vortices resulting from the smooth assembling of structures shed from the initially steady flow patterns.
This scenario has never been exposed so far and differs from those given in previous works \citep{hy04,sa83,yik07,wz09}.
To partly explain this difference, we have stressed the effects of the vertical flow confinement which enhances the base vortices and induces a more efficient entrainment of the head of the hairpin vortices.
The latter is therefore located at the downstream end of the hairpin and the legs are almost parallel to the streamwise axis.
Also, we have detected the formation of a chain of $\Omega$-shaped vortices as a consequence of the streamwise orientation of legs of the hairpin vortices.
The $\Omega$-shaped vortices are well known in turbulent boundary layers and their formation stems from the same mechanism as in this classic case.
For $Re\ge250$, the appearance and enhancement of vertical oscillations of the base vortices causes the vortex street to turn asymmetric and irregular, while hairpin vortices become chaotic aggregates of structures.

We have reported the evolutions of a set of flow coefficients with $Re$.
Firstly, the spanwise lift coefficient has been found to increase when the recirculation appeared on the top face.
Secondly, the transition to asymmetry in the unsteady wake has been shown to induce a collapse in the Strouhal number that reflected the disappearance of the symmetric mode associated with the hairpin vortex shedding we have identified.

Finally, although we have clearly identified the dynamics of the hairpin vortices, questions about the respective influences of the various parameters set in our numerical computations are still open.
Firstly, the influence of the cylinder aspect ratio $\gamma$ would require a parametric investigation.
Although it is well accepted that $\gamma$ determines whether the onset of vortex shedding gives rise to hairpin or K\'arm\'an vortices, the effect of $\gamma$ within any of these regimes is yet to be clarified.
Secondly, we have used physical walls at the outer boundaries of our domain.
Though physically realistic, both the transverse and spanwise confinements should be subject to a thorough scrutiny to determine their effects on the stability and the shape of the hairpin vortices.
The transverse confinement in the present configuration is low enough to neglect its effects.
One may however expect valuable information on whether moderate transverse blockage improves the stability of the hairpin vortices.
By contrast, we have shown that the spanwise blockage had a significant influence on the flow dynamics.
One could expect a lighter spanwise confinement to minimise the role of the base vortices in these dynamics and result in a modification of the overall shape of the hairpin vortices which could be closer to the arch-type vortices observed in experiments by \cite{sa83} and \cite{wz09}.

Supplementary movie available at journals.cambridge.org/flm

%%%%%%%%%%%%%%%%%%%%%%%%%%%%%%%%%%%%%%%%%%%%%%%%%%
%                                                %
%               Bibliography                     %
%                                                %
%%%%%%%%%%%%%%%%%%%%%%%%%%%%%%%%%%%%%%%%%%%%%%%%%%

\end{document}